\documentclass[fleqn,10pt]{wlscirep}
\usepackage[utf8]{inputenc}
\usepackage[T1]{fontenc}
\usepackage{tabularx}
\usepackage{threeparttable}
\usepackage{pifont}
\usepackage{siunitx}
\usepackage{amsmath}
\usepackage{graphicx}
\usepackage{booktabs}
\usepackage{array}
\usepackage{float}
\usepackage{url}
\newcommand{\cmark}{\ding{51}}

\title{Noninvasive H3 K27M screening in pediatric diffuse midline glioma using radiomics on heterogeneous T2-weighted MRI}

\author[1,*]{Arthur Zagitov}
\author[2]{Alexander Beznosikov}
\author[3]{Vladimir Bozhenko}
\author[1]{Ninel Kamyshnikova}
\author[3]{Tatiana Kulinich}
\author[1]{Sofia Polozova}
\author[1,4]{Yaroslav Kholodov}

\affil[1]{Innopolis University, Innopolis, Russia}
\affil[2]{BRAIn Lab, Moscow, Russia}
\affil[3]{Russian Research Center of Roentgenology and Radiology, Moscow, Russia}
\affil[4]{Moscow Institute of Physics and Technology, Moscow, Russia}
\affil[*]{artazagitov@gmail.com}


\begin{abstract}
Histone H3K27M mutation status defines a clinically aggressive subgroup of pediatric diffuse midline glioma and informs prognosis and trial eligibility, but confirmation usually requires tissue sampling from eloquent midline structures. We evaluated whether radiomics from routinely available T2-weighted MRI can provide an adjunctive screening signal in a heterogeneous referral-style cohort, where scans are often acquired externally and T2-weighted imaging is the only consistently available sequence. Ninety-eight pediatric patients with tissue-confirmed status were analyzed (73 mutation-positive, 25 wild-type). Expert tumor segmentations defined the regions of interest for PyRadiomics feature extraction after isotropic resampling, dual skull stripping, and multi-scale filtering. We systematically ablated preprocessing, correlation pruning with repeated recursive feature elimination, tumor volume, and TabDDPM synthetic minority augmentation across 100 stratified train/test splits with real-only test sets. Pure radiomics achieved accuracy 0.664 and F1-score 0.784. The best pipeline used preprocessing, feature selection, and volume with CatBoost, achieving accuracy 0.730$\pm$0.068 and F1-score 0.826$\pm$0.044. TabDDPM improved TabPFN to F1-score 0.81$\pm$0.05 at 200 augmented rows. These results support T2-weighted radiomics as a moderate screening and triage aid, not a replacement for tissue-based diagnosis.
\end{abstract}
\begin{document}

\raggedbottom
\maketitle
\thispagestyle{empty}

\section*{Introduction}
Diffuse midline glioma is an aggressive pediatric central nervous system tumor with limited therapeutic options and poor survival. Risk stratification and clinical trial eligibility increasingly depend on molecular alterations, including the histone H3K27M mutation, which is associated with adverse prognosis and influences management decisions\cite{intro1}. Mutation status is usually established using tissue sampling followed by histologic and molecular assays\cite{intro2}. However, diffuse midline gliomas often arise in eloquent midline structures, where biopsy can be technically challenging and may carry neurological risk. In addition, small tissue samples may not fully capture intratumoral heterogeneity\cite{intro3}. These factors create a need for imaging-based tools that can support, rather than replace, molecular confirmation.

Magnetic resonance imaging is central to diagnosis, treatment planning, and follow-up. Conventional interpretation describes tumor location, signal, and enhancement patterns, but these qualitative features do not directly encode molecular genotype. Radiomics converts medical images into quantitative descriptors of shape, intensity distribution, and spatial texture\cite{rew_radiomics1,intro4,rew_radiomics3}. Because radiomic features can be affected by acquisition settings and reconstruction differences, reproducibility and standardization are central concerns\cite{rew_radiomics5,rew_radiomics6}. Radiomics has been increasingly used across oncologic imaging tasks, including clinically relevant classification and prognostic applications\cite{rew_radiomics2}. This issue is especially relevant for pediatric diffuse midline glioma because cohorts are small, class imbalance is common, and scans may come from multiple scanners and institutions.

Prior studies have reported H3K27M prediction using radiomics, automated machine learning, multiparametric magnetic resonance imaging, and deep learning\cite{rew_prior1,rew_prior3,rew_prior2}. Several barriers remain for translation. First, pediatric diffuse midline glioma is rare, and mutation-confirmed cohorts are typically small and imbalanced. Second, models developed in controlled imaging cohorts may not perform as well when applied to scans acquired across different vendors, institutions, and clinical protocols. Third, multiparametric approaches can exploit additional biological contrast, but they require sequences that are not consistently available in externally referred patients. Finally, many prior reports emphasize a single final pipeline rather than isolating the contribution of preprocessing, feature selection, volume, or data-balancing strategy.

The clinical workflow motivating this study is a referral scenario in which children often arrive at a specialized center with magnetic resonance imaging already performed elsewhere. In this setting, axial T2-weighted imaging is frequently the only sequence available in a diagnostic-quality and comparable form across patients. A decision-support method intended for this workflow should therefore be able to operate on single-sequence data of variable provenance rather than on a curated multiparametric protocol. We deliberately designed the study around this constraint. Heterogeneous acquisition and T2-only input were treated as defining properties of the intended use case, not as nuisance factors to be removed. The goal was correspondingly modest: to provide a low-cost signal for prioritizing molecular confirmation, trial screening, or expert review, while recognizing that tissue-based diagnosis remains the reference standard.

We evaluated an ablation-driven radiomics framework using expert tumor segmentations from T2-weighted magnetic resonance imaging. The analysis quantified the contributions of preprocessing intended to mitigate scanner and protocol variability\cite{rew_radiomics6,demirciouglu2022effect,park2021robustness,thakur2020brain}, correlation-based and stability-oriented feature selection adapted to high-dimensional radiomics\cite{feature_selection}, explicit tumor volume, and diffusion-based tabular synthesis with TabDDPM as an alternative to heuristic oversampling\cite{tabddpm,synth_2}. We benchmarked TabPFN, a strong small-data tabular learner, alongside established ensemble, boosting, linear, and distance-based classifiers\cite{tabpfn}.

\section*{Results}

\subsection*{Cohort and imaging setting}
\begin{figure}[H]
\centering
\includegraphics[width=\linewidth]{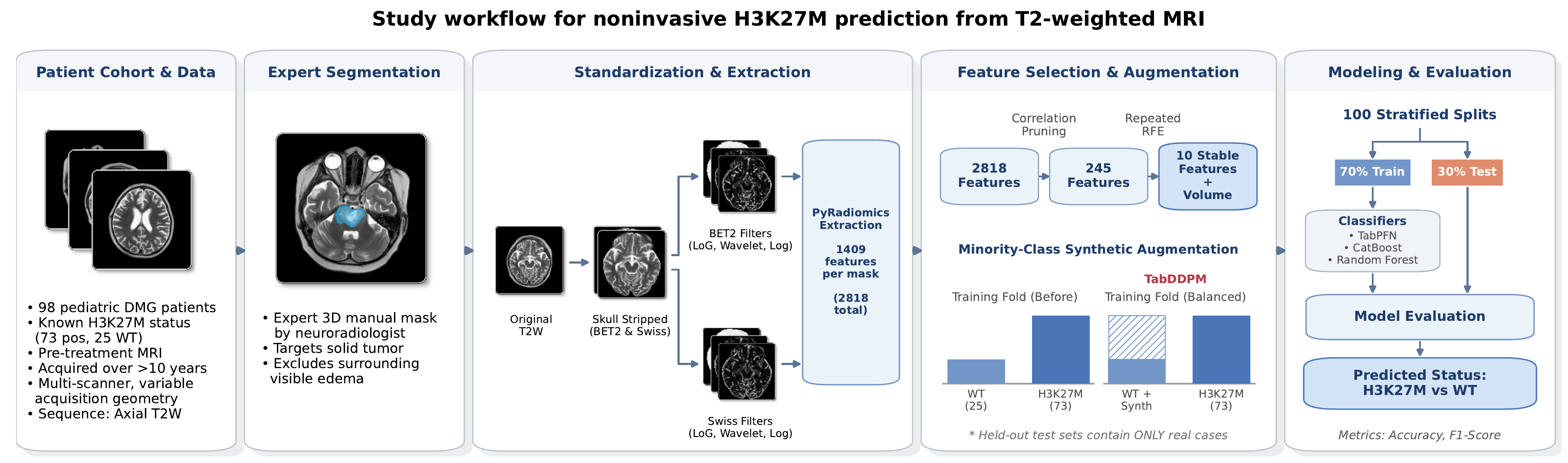}
\caption{Study workflow for H3K27M screening from T2-weighted MRI. The workflow included patient selection, expert tumor segmentation, skull stripping, filtered-image generation, PyRadiomics feature extraction, correlation pruning, repeated recursive feature elimination, optional TabDDPM synthetic minority augmentation, and classifier evaluation. Synthetic samples were generated within training folds only; test folds contained real patients only.}
\label{fig:workflow}
\end{figure}

The cohort comprised 98 pediatric patients with diffuse midline glioma and known H3K27M status. Seventy-three patients were mutation-positive and 25 were wild-type. All patients had pre-treatment T2-weighted magnetic resonance imaging and expert tumor segmentation masks. The scans were collected over more than 10 years and originated from multiple institutions, scanners, and clinical protocols, with variable voxel geometry. This heterogeneity was deliberately retained to reflect the referral scenario in which treatment teams may receive scans acquired at external sites. The workflow from patient selection through segmentation, feature extraction, dimensionality reduction, training-only synthetic augmentation, and repeated split evaluation is summarized in Fig.~\ref{fig:workflow}.

\subsection*{Preprocessing, feature selection, and volume improved the strongest pipelines}
\begin{figure}[H]
\centering
\includegraphics[width=\linewidth]{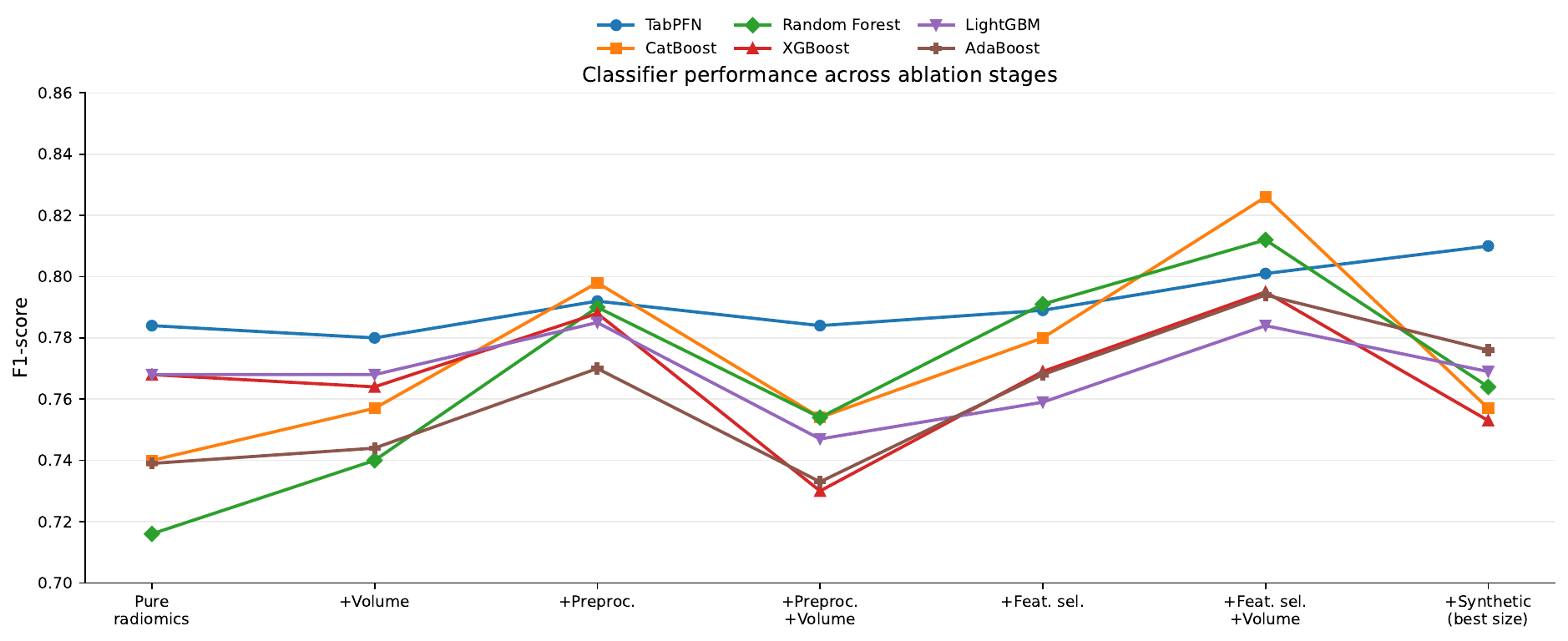}
\caption{Classifier performance across ablation stages. Points show mean F1-score across 100 stratified 70/30 train/test splits. The final synthetic stage corresponds to the best-performing TabDDPM-augmented training-table size for each classifier. Preproc., preprocessing; Feat. sel., feature selection.}
\label{fig:ablations}
\end{figure}

Pure radiomics from the original T2-weighted images achieved accuracy 0.664 and F1-score 0.784 with TabPFN. After preprocessing, correlation pruning, repeated recursive feature elimination, and tumor volume inclusion, the best overall performance was obtained with CatBoost, reaching accuracy 0.730$\pm$0.068 and F1-score 0.826$\pm$0.044 (Table~\ref{tab:best-results}). Random forest and TabPFN also achieved F1-scores near 0.81 in their best configurations. Across ablation stages, preprocessing and feature selection were the most consistent contributors to improved F1-score, whereas volume provided complementary signal mainly after feature selection (Fig.~\ref{fig:ablations}). The strongest configurations combined high recall for the mutation-positive class with moderate precision, an error profile consistent with a screening role in which missing mutation-positive patients is more problematic than over-referral for confirmatory testing.

\begin{table}[H]
\centering
\caption{Best observed results for each classifier and the corresponding pipeline components. A checkmark indicates that the component was used. Metrics are mean$\pm$standard deviation over 100 random splits.}
\label{tab:best-results}
\footnotesize
\setlength{\tabcolsep}{2.6pt}
\renewcommand{\arraystretch}{1.12}
\begin{tabular}{@{}>{\raggedright\arraybackslash}p{0.19\linewidth}
*{4}{>{\centering\arraybackslash}p{0.036\linewidth}}
*{4}{>{\centering\arraybackslash}p{0.125\linewidth}}@{}}
\toprule
\textbf{Model} & \textbf{Pre.} & \textbf{Sel.} & \textbf{Vol.} & \textbf{Syn.} & \textbf{Acc.} & \textbf{Recall} & \textbf{Prec.} & \textbf{F1} \\
\midrule
TabPFN & \cmark & \cmark & \cmark & \cmark & 0.71$\pm$0.10 & 0.85$\pm$0.05 & 0.78$\pm$0.09 & 0.81$\pm$0.05 \\
Adaptive Boosting & \cmark & \cmark & \cmark & -- & 0.69$\pm$0.11 & 0.81$\pm$0.11 & 0.79$\pm$0.06 & 0.79$\pm$0.08 \\
Logistic Regression & \cmark & -- & -- & -- & 0.67$\pm$0.10 & 0.72$\pm$0.12 & 0.82$\pm$0.08 & 0.76$\pm$0.09 \\
LightGBM & -- & -- & \cmark & -- & 0.68$\pm$0.10 & 0.78$\pm$0.13 & 0.77$\pm$0.07 & 0.77$\pm$0.08 \\
KNN & \cmark & \cmark & \cmark & -- & 0.63$\pm$0.08 & 0.69$\pm$0.10 & 0.80$\pm$0.07 & 0.73$\pm$0.07 \\
SVM & \cmark & \cmark & \cmark & \cmark & 0.68$\pm$0.08 & 0.99$\pm$0.03 & 0.68$\pm$0.08 & 0.80$\pm$0.06 \\
Random Forest & \cmark & \cmark & \cmark & -- & 0.71$\pm$0.08 & 0.84$\pm$0.08 & 0.79$\pm$0.05 & 0.81$\pm$0.06 \\
Decision Tree & \cmark & \cmark & \cmark & \cmark & 0.66$\pm$0.11 & 0.73$\pm$0.10 & 0.76$\pm$0.13 & 0.74$\pm$0.10 \\
Ridge Regression & \cmark & \cmark & \cmark & \cmark & 0.65$\pm$0.12 & 0.74$\pm$0.14 & 0.75$\pm$0.11 & 0.73$\pm$0.11 \\
AutoML & \cmark & \cmark & \cmark & -- & 0.70$\pm$0.13 & 0.83$\pm$0.09 & 0.79$\pm$0.14 & 0.80$\pm$0.09 \\
CatBoost & \cmark & \cmark & \cmark & -- & \textbf{0.73$\pm$0.07} & 0.85$\pm$0.06 & 0.80$\pm$0.05 & \textbf{0.83$\pm$0.04} \\
XGBoost & \cmark & \cmark & \cmark & -- & 0.69$\pm$0.11 & 0.81$\pm$0.09 & 0.79$\pm$0.07 & 0.80$\pm$0.08 \\
\bottomrule
\end{tabular}
\vspace{2pt}
\begin{minipage}{\linewidth}
\footnotesize Pre., preprocessing; Sel., feature selection; Vol., tumor volume; Syn., TabDDPM synthetic augmentation. Preprocessing includes isotropic resampling, dual skull stripping, and filtered-image feature maps. Feature selection includes correlation pruning and repeated recursive feature elimination.
\end{minipage}
\end{table}

\subsection*{Synthetic minority augmentation had classifier-specific benefit}
TabDDPM was evaluated as a training-only augmentation method for the minority wild-type class. In an auxiliary comparison using a random forest downstream classifier, TabDDPM outperformed no augmentation and SMOTE, reaching F1-score 0.71 compared with 0.60 and 0.66, respectively (Table~\ref{tab:augmentation}). In the full ablation study, the clearest benefit was observed for TabPFN, which improved from F1-score 0.80$\pm$0.06 without synthetic samples to 0.81$\pm$0.05 at 200 augmented rows. Larger augmented training-table sizes did not consistently improve performance, and other classifiers showed limited or no benefit (Fig.~\ref{fig:synth}). Complete per-model results for all augmentation sizes are provided in Supplementary Tables~S8-S13.

\begin{table}[H]
\centering
\caption{Synthetic minority augmentation methods evaluated with a random forest classifier on a held-out real-patient test fold.}
\label{tab:augmentation}
\begin{tabular}{lccc}
\toprule
\textbf{Method} & \textbf{F1-score} & \textbf{Precision} & \textbf{Recall} \\
\midrule
No augmentation & 0.60 & 0.70 & 0.56 \\
SMOTE & 0.66 & 0.67 & 0.65 \\
TabDDPM & 0.71 & 0.72 & 0.71 \\
\bottomrule
\end{tabular}
\end{table}

\begin{figure}[H]
\centering
\includegraphics[width=0.72\linewidth]{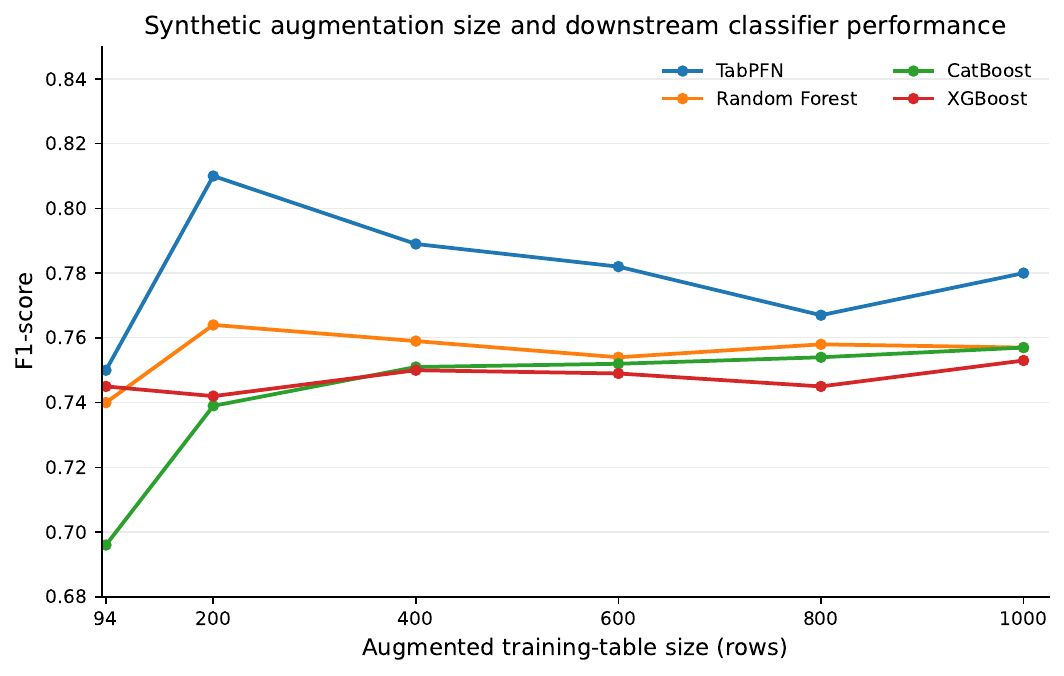}
\caption{Effect of augmented training-table size on downstream classifier performance. F1-score is shown after feature selection and inclusion of tumor volume. Synthetic samples were added only to training folds.}
\label{fig:synth}
\end{figure}

\section*{Discussion}
Radiomics from heterogeneous clinical T2-weighted magnetic resonance imaging provided moderate prediction of H3K27M mutation status in pediatric diffuse midline glioma. The central contribution is not a claim of biopsy replacement, but a practical demonstration that a single-sequence, referral-style imaging workflow can provide an adjunctive screening signal. The most consistent improvements were associated with preprocessing that standardized voxel geometry and enabled multi-scale feature extraction, and with redundancy reduction through correlation pruning followed by stability-oriented recursive feature elimination. After feature selection, explicit inclusion of tumor volume added complementary information and improved ensemble methods. Diffusion-based augmentation was not uniformly beneficial across classifiers, but it improved TabPFN and yielded its best overall configuration.

The achieved performance should be interpreted in light of the intended-use setting. Several radiomics and deep-learning studies have reported higher headline metrics for H3K27M prediction, but many used multiparametric imaging or more controlled acquisition protocols. Guo et al. compared multiple sequence combinations and machine learning methods in a pathologically confirmed cohort and reported that the best multiparametric combination reached an area under the receiver operating characteristic curve of 0.969, while performance varied substantially across sequences and classifiers\cite{guo2022}. Such results demonstrate the potential value of richer imaging, but they are not directly comparable to a single-sequence T2-weighted workflow. Metrics also differ across studies; area under the receiver operating characteristic curve, accuracy, and F1-score summarize different aspects of performance.

A more comparable external-validation context suggests that performance near the low-to-mid 70\% accuracy range is plausible when conventional T2-weighted radiomics is tested across institutions. Indoria et al. reported approximately 80\% internal accuracy for several classifiers, but external test accuracy decreased to 75\% for decision tree, 73\% for random forest, and 65.1\% for k-nearest neighbors\cite{indoria2024}. Our best accuracy of 0.73 in a deliberately heterogeneous referral-style cohort is consistent with that range. This does not prove external validity for our model, but it contextualizes why performance is lower than highly optimized multiparametric reports. It also supports the idea that acquisition heterogeneity and single-sequence input are meaningful stress tests rather than merely methodological shortcomings.

The appropriate clinical role follows from this performance profile. The observed accuracy and F1-score are not diagnostic-grade and do not support use of the model as a substitute for biopsy, immunohistochemistry, sequencing, or other molecular assays. A more realistic role is triage within a human-in-the-loop workflow: flagging patients for prioritized molecular confirmation, supporting trial prescreening, and providing a complementary signal when tissue is delayed, limited, or not feasible to obtain. In the strongest configurations, recall for the H3K27M-positive class was high, including 0.85 for CatBoost and TabPFN and 0.99 for SVM at lower precision. This error profile may be useful for prioritization because missing a mutation-positive case is usually more consequential than sending a wild-type case for additional confirmatory assessment. The treating clinician would retain decision authority, and the model would contribute one input among imaging, clinical, anatomic, and institutional considerations.

The technical findings are also relevant for future radiomics work in small pediatric cohorts. Preprocessing and multi-scale feature extraction were beneficial in the presence of variable acquisition geometry. Feature selection was important because the preprocessed feature table contained 2818 radiomic features for only 98 patients, a setting with substantial overfitting risk. Tumor volume became useful mainly after redundancy reduction, suggesting that an interpretable anatomic descriptor can complement texture features when high-dimensional collinearity is controlled. Synthetic augmentation required careful evaluation rather than automatic adoption. TabDDPM improved TabPFN at a moderate augmentation size, but larger synthetic tables did not reliably improve performance, and most other classifiers showed limited benefit.

This study has limitations. It is retrospective and single-center in the sense that model development and evaluation were performed within one institutional repository, although the imaging data span many external acquisition sources. Evaluation used repeated random splits rather than a fully independent external test set. Intentional acquisition heterogeneity probes robustness, but it is not a substitute for true external validation, and overlapping splits yield descriptive summaries of split-to-split variability rather than independent inferential estimates. Complete scanner and protocol metadata were unavailable for many externally acquired examinations, preventing subgroup analysis by vendor, field strength, or sequence parameters. Only T2-weighted imaging was analyzed; this was intentional for practical applicability, but multiparametric magnetic resonance imaging may improve performance when complete protocols are available. The cohort is modest in size, reflecting the rarity of mutation-confirmed pediatric diffuse midline glioma, and this precluded some subgroup analyses. An additional cohort without molecular ground truth could not be incorporated into the primary analysis; exploratory pseudo-labeling from event-free survival and unsupervised radiomics grouping did not improve prediction, suggesting that survival alone is an imperfect surrogate for mutation status. Finally, synthetic augmentation was performed in radiomic feature space and may not capture image-domain variability.

In summary, a radiomics and machine-learning framework applied to heterogeneous clinical T2-weighted magnetic resonance imaging enables moderate noninvasive estimation of H3K27M status in pediatric diffuse midline glioma. The results support a decision-support and triage role for prioritizing molecular confirmation and clinical trial screening, not replacement of tissue-based diagnosis. Preprocessing and feature selection improved performance and stability, tumor volume added complementary signal after selection, and diffusion-based tabular augmentation improved selected small-data classifiers.

\section*{Methods}

\subsection*{Ethics and consent}
This retrospective study was approved by the local institutional ethics committee. All procedures were performed in accordance with relevant guidelines and regulations. Informed consent was waived by the committee because the study used anonymized retrospective imaging and clinical data. Study participant names and personally identifiable information were removed from all text, figures, tables, and images.

\subsection*{Study design and patients}
Consecutive pediatric patients younger than 18 years with a diagnosis of diffuse midline glioma and known H3K27M status were identified from a clinical repository. Inclusion criteria were availability of pre-treatment magnetic resonance imaging including a T2-weighted sequence, tissue-based determination of H3K27M status, and an expert tumor segmentation mask. Exclusion criteria were post-treatment imaging only, missing segmentation, or non-diagnostic image quality. An exploratory set of 125 pediatric diffuse midline glioma cases without known H3K27M status was available but excluded from the primary analysis because molecular ground truth was absent.

\subsection*{MRI selection and acquisition heterogeneity}
Examinations were collected over more than 10 years and originated from multiple institutions, scanners, and clinical protocols, resulting in substantial variability in voxel geometry, slice sampling, and acquisition parameters. Complete scanner and sequence metadata were not consistently available because many examinations were acquired before referral and archived with variable metadata completeness. For standardization and maximal cohort size, the axial T2-weighted sequence was selected for radiomic analysis when multiple sequences were present. In many referred studies, T2-weighted imaging was the only diagnostic-quality sequence available. This choice reflects the intended deployment scenario rather than a data-quality compromise: a model restricted to a single, near-universally available sequence can be applied to the broadest possible referral population, whereas a multiparametric requirement would exclude patients without complete protocols.

\subsection*{Tumor segmentation and region of interest definition}
Tumor regions of interest were manually delineated on axial T2-weighted images in 3D Slicer v3.5.2\cite{slicer} by a neuroradiologist with 7 years of experience. Segmentations targeted the visually apparent solid tumor component on T2-weighted imaging and avoided surrounding visible edema when distinguishable. These masks served as the region of interest for all downstream radiomic computations.

\subsection*{Preprocessing and derived image generation}
Preprocessing was selected to reduce acquisition-driven variability and enable multi-scale radiomic characterization\cite{rew_radiomics6,demirciouglu2022effect}. T2-weighted images and masks were resampled to isotropic $1\times1\times1$~mm$^3$ resolution using trilinear interpolation for images and nearest-neighbor interpolation for masks, which reduces voxel-geometry bias in texture features\cite{park2021robustness}. Two skull-stripping outputs were generated with BET2\cite{bet2} and SwissSkullStripper in 3D Slicer\cite{slicer}. Both outputs were propagated through feature extraction to reduce sensitivity to failure modes of any single skull-stripping approach, which can be imperfect in tumor cases\cite{thakur2020brain}. In addition to the original skull-stripped image, derived image maps included wavelet decompositions, Laplacian-of-Gaussian filtering with $\sigma=1.0$ and $3.0$, and nonlinear intensity transforms (logarithm, exponential, square, and square root).

\subsection*{Radiomic feature extraction}
Radiomic features were extracted in 3D using PyRadiomics\cite{pyRadiomics}, a Python package for standardized radiomic feature extraction compliant with the Image Biomarker Standardization Initiative\cite{rew_radiomics5}. Feature classes included shape, first-order statistics, gray-level co-occurrence matrix, gray-level run length matrix, gray-level size zone matrix, gray-level dependence matrix, and neighborhood gray tone difference matrix features (Table~\ref{tab:feature-classes}). Baseline unfiltered radiomics yielded 107 features per case. After filtering, features were extracted from the original image plus 14 derived images (8 wavelet sub-bands, 2 Laplacian-of-Gaussian scales, and 4 nonlinear transforms). Shape features are not filter-dependent and were computed once. This yielded 1409 features per skull-stripping method, giving 2818 total radiomic features per patient.

\begin{table}[H]
\centering
\caption{Radiomic feature classes and phenotypic interpretations.}
\label{tab:feature-classes}
\begin{tabularx}{\linewidth}{@{}>{\raggedright\arraybackslash}p{0.27\linewidth}X@{}}
\toprule
\textbf{Feature class} & \textbf{Interpretation} \\
\midrule
Shape & Size, geometric shape, and surface characteristics of the tumor region. \\
First-order statistics & Global distribution of voxel intensities within the tumor mask. \\
GLCM & Spatial relationships between adjacent voxel intensities and fine texture. \\
GLRLM & Consecutive runs of voxels with identical intensities, reflecting coarse or fine texture. \\
GLSZM & Connected regions sharing identical intensity values, reflecting regional homogeneity. \\
GLDM & Dependence between a center voxel and neighboring voxels, reflecting local texture coarseness. \\
NGTDM & Difference between a voxel and its immediate neighborhood, reflecting spatial contrast. \\
\bottomrule
\end{tabularx}
\end{table}

\subsection*{Tumor volume}
Tumor volume was included as an explicit scalar descriptor because tumor burden and growth patterns may differ by biology and are clinically interpretable. It was computed from the resampled tumor mask as
\begin{equation}
V = N_{\mathrm{vox}} \cdot s_x s_y s_z,
\end{equation}
where $N_{\mathrm{vox}}$ is the number of tumor voxels and $(s_x,s_y,s_z)$ are voxel spacings after resampling. Because the pipeline resampled to 1~mm isotropic voxels, each voxel contributed 1~mm$^3$.

\subsection*{Feature selection}
Feature selection used a two-stage approach adapted to high-dimensional radiomics in small cohorts\cite{feature_selection}. First, duplicate columns were removed and correlation pruning was applied with Pearson $|r|>0.8$. Features were grouped into shape, first-order, and texture categories. Within each group, the feature with higher average absolute correlation was removed from highly correlated pairs. Between-group correlations were then reduced using a hierarchical removal order from texture to first-order to shape, prioritizing lower-order interpretable descriptors where redundant. Second, recursive feature elimination with a random forest estimator was repeated across 100 training splits, and the 10 most consistently high-ranked features were retained. The selected features and ranking statistics are provided in Supplementary Table~S5.

\subsection*{Synthetic augmentation}
TabDDPM, a diffusion model for tabular data\cite{tabddpm,synth_2}, was trained on the selected 10-feature radiomic subset within each training fold only. Conditional generation used two classes, H3K27M-positive and wild-type. Synthetic samples were generated for the minority wild-type class to create balanced training datasets. Six augmented training-table sizes were evaluated: 94, 200, 400, 600, 800, and 1000 total rows. Test sets were not augmented and contained only real patients. TabDDPM hyperparameters are reported in Supplementary Tables~S14 and S15. In an auxiliary comparison, TabDDPM was compared with SMOTE\cite{smote} using a random forest classifier trained on augmented training folds and evaluated on untouched real test folds.

\subsection*{Classifiers and evaluation}
Classifiers included TabPFN\cite{tabpfn}, LightGBM, XGBoost, CatBoost, random forest, adaptive boosting, decision tree, logistic regression, ridge logistic regression, support vector machine, k-nearest neighbors, and AutoML. Evaluation used 100 repeated stratified random 70/30 train/test splits. Feature selection and synthetic generation were performed within each training split to avoid leakage. Test folds contained only real patients. H3K27M-positive status was treated as the positive class. Accuracy, recall, precision, and F1-score were summarized as mean$\pm$standard deviation across 100 splits. The F1-score was computed as
\begin{equation}
\mathrm{F1} = \frac{2\,\mathrm{Precision}\cdot\mathrm{Recall}}{\mathrm{Precision}+\mathrm{Recall}}.
\end{equation}
The analysis was descriptive. Because the 100 splits were drawn from the same cohort and overlapped, these summaries characterize split-to-split variability rather than independent samples. Reporting was mapped to the TRIPOD+AI checklist for clinical prediction model studies that use regression or machine-learning methods; the item-level reporting map is provided in Supplementary Table~S16\cite{tripodai}.

\subsection*{Manuscript preparation}
Large language model assistance was used for language editing and formatting. The authors are responsible for all scientific content, analyses, interpretations, and final text.

\section*{Data Availability}
A code repository containing preprocessing scripts and model-training code is available at \url{https://github.com/threeteck/H3K27M_Radiomics}. The raw imaging data and patient-level clinical data cannot be publicly released because they contain sensitive pediatric medical information and are subject to institutional and data-use restrictions. Requests for controlled access to de-identified derived radiomic features may be directed to the corresponding author and will be considered subject to approval by the data-providing institution.

\bibliography{refs}

@article{intro1,
  title={Targeted detection of genetic alterations reveal the prognostic impact of {H3K27M} and {MAPK} pathway aberrations in paediatric thalamic glioma},
  author={Ryall, Scott and Krishnatry, Rahul and Arnoldo, Anthony and Buczkowicz, Pawel and Mistry, Matthew and Siddaway, Robert and Ling, Cino and Pajovic, Sanja and Yu, Man and Rubin, Joshua B. and others},
  journal={Acta Neuropathologica Communications},
  volume={4},
  pages={1--10},
  year={2016},
  publisher={Springer}
}

@article{intro2,
  title={Brainstem biopsy in pediatric diffuse intrinsic pontine glioma in the era of precision medicine: the {INFORM} study experience},
  author={Pfaff, Elke and El Damaty, Ahmed and Balasubramanian, Gnana Prakash and Blattner-Johnson, Mirjam and Worst, Barbara C. and Stark, Sebastian and Witt, Hendrik and Pajtler, Kristian W. and van Tilburg, Cornelis M. and Witt, Ruth and others},
  journal={European Journal of Cancer},
  volume={114},
  pages={27--35},
  year={2019},
  publisher={Elsevier}
}

@article{intro3,
  title={Diffuse glioma heterogeneity and its therapeutic implications},
  author={Nicholson, James G. and Fine, Howard A.},
  journal={Cancer Discovery},
  volume={11},
  number={3},
  pages={575--590},
  year={2021},
  publisher={AACR}
}

@article{rew_radiomics1,
  title={Radiomics: extracting more information from medical images using advanced feature analysis},
  author={Lambin, Philippe and Rios-Velazquez, Emmanuel and Leijenaar, Ralph and Carvalho, Sara and Van Stiphout, Ruud G. P. M. and Granton, Patrick and Zegers, Catharina M. L. and Gillies, Robert and Boellard, Ronald and Dekker, Andr{\'e} and others},
  journal={European Journal of Cancer},
  volume={48},
  number={4},
  pages={441--446},
  year={2012},
  publisher={Elsevier}
}

@article{intro4,
  title={Radiomics: the bridge between medical imaging and personalized medicine},
  author={Lambin, Philippe and Leijenaar, Ralph T. H. and Deist, Timo M. and Peerlings, Jurgen and De Jong, Evelyn E. C. and Van Timmeren, Janita and Sanduleanu, Sebastian and Larue, Ruben T. H. M. and Even, Aniek J. G. and Jochems, Arthur and others},
  journal={Nature Reviews Clinical Oncology},
  volume={14},
  number={12},
  pages={749--762},
  year={2017},
  publisher={Nature Publishing Group}
}

@article{rew_radiomics3,
  title={The potential of radiomic-based phenotyping in precision medicine: a review},
  author={Aerts, Hugo J. W. L.},
  journal={JAMA Oncology},
  volume={2},
  number={12},
  pages={1636--1642},
  year={2016},
  publisher={American Medical Association}
}

@article{rew_radiomics5,
  title={The image biomarker standardization initiative: standardized quantitative radiomics for high-throughput image-based phenotyping},
  author={Zwanenburg, Alex and Valli{\`e}res, Martin and Abdalah, Mahmoud A. and Aerts, Hugo J. W. L. and Andrearczyk, Vincent and Apte, Aditya and Ashrafinia, Saeed and Bakas, Spyridon and Beukinga, Roelof J. and Boellaard, Ronald and others},
  journal={Radiology},
  volume={295},
  number={2},
  pages={328--338},
  year={2020},
  publisher={Radiological Society of North America}
}

@article{rew_radiomics6,
  title={Radiomics in oncology: a practical guide},
  author={Shur, Joshua D. and Doran, Simon J. and Kumar, Santosh and Ap Dafydd, Derfel and Downey, Kate and O'Connor, James P. B. and Papanikolaou, Nikolaos and Messiou, Christina and Koh, Dow-Mu and Orton, Matthew R.},
  journal={RadioGraphics},
  volume={41},
  number={6},
  pages={1717--1732},
  year={2021},
  publisher={Radiological Society of North America}
}

@article{rew_radiomics2,
  title={A review of original articles published in the emerging field of radiomics},
  author={Song, Jiangdian and Yin, Yanjie and Wang, Hairui and Chang, Zhihui and Liu, Zhaoyu and Cui, Lei},
  journal={European Journal of Radiology},
  volume={127},
  pages={108991},
  year={2020},
  publisher={Elsevier}
}

@article{rew_prior1,
  title={Automated machine learning based on radiomics features predicts {H3 K27M} mutation in midline gliomas of the brain},
  author={Su, Xiaorui and Chen, Ni and Sun, Huaiqiang and Liu, Yanhui and Yang, Xibiao and Wang, Weina and Zhang, Simin and Tan, Qiaoyue and Su, Jingkai and Gong, Qiyong and others},
  journal={Neuro-Oncology},
  volume={22},
  number={3},
  pages={393--401},
  year={2020},
  publisher={Oxford University Press}
}

@article{rew_prior3,
  title={Machine learning-based multiparametric magnetic resonance imaging radiomics for prediction of {H3K27M} mutation in midline gliomas},
  author={Kandemirli, Sedat Giray and Kocak, Burak and Naganawa, Shotaro and Ozturk, Kerem and Yip, Stephen S. F. and Chopra, Saurav and Rivetti, Luciano and Aldine, Amro Saad and Jones, Karra and Cayci, Zuzan and others},
  journal={World Neurosurgery},
  volume={151},
  pages={e78--e85},
  year={2021},
  publisher={Elsevier}
}

@article{rew_prior2,
  title={Deep learning-based prediction of {H3K27M} alteration in diffuse midline gliomas based on whole-brain {MRI}},
  author={Huang, Bowen and Zhang, Yuekang and Mao, Qing and Ju, Yan and Liu, Yanhui and Su, Zhengzheng and Lei, Yinjie and Ren, Yanming},
  journal={Cancer Medicine},
  volume={12},
  number={16},
  pages={17139--17148},
  year={2023},
  publisher={Wiley Online Library}
}

@article{demirciouglu2022effect,
  title={The effect of preprocessing filters on predictive performance in radiomics},
  author={Demircio{\u{g}}lu, Aydin},
  journal={European Radiology Experimental},
  volume={6},
  number={1},
  pages={40},
  year={2022},
  publisher={Springer}
}

@article{park2021robustness,
  title={Robustness of magnetic resonance radiomic features to pixel size resampling and interpolation in patients with cervical cancer},
  author={Park, Shin-Hyung and Lim, Hyejin and Bae, Bong Kyung and Hahm, Myong Hun and Chong, Gun Oh and Jeong, Shin Young and Kim, Jae-Chul},
  journal={Cancer Imaging},
  volume={21},
  pages={1--11},
  year={2021},
  publisher={Springer}
}

@article{thakur2020brain,
  title={Brain extraction on {MRI} scans in presence of diffuse glioma: multi-institutional performance evaluation of deep learning methods and robust modality-agnostic training},
  author={Thakur, Siddhesh and Doshi, Jimit and Pati, Sarthak and Rathore, Saima and Sako, Chiharu and Bilello, Michel and Ha, Sung Min and Shukla, Gaurav and Flanders, Adam and Kotrotsou, Aikaterini and others},
  journal={NeuroImage},
  volume={220},
  pages={117081},
  year={2020},
  publisher={Elsevier}
}

@article{feature_selection,
  title={{``Real-world''} radiomics from multi-vendor {MRI}: an original retrospective study on the prediction of nodal status and disease survival in breast cancer, as an exemplar to promote discussion of the wider issues},
  author={Doran, Simon J. and Kumar, Santosh and Orton, Matthew and d'Arcy, James and Kwaks, Fenna and O'Flynn, Elizabeth and Ahmed, Zaki and Downey, Kate and Dowsett, Mitch and Turner, Nicholas and others},
  journal={Cancer Imaging},
  volume={21},
  number={1},
  pages={37},
  year={2021},
  publisher={Springer}
}

@inproceedings{tabddpm,
  title={{TabDDPM}: modelling tabular data with diffusion models},
  author={Kotelnikov, Akim and Baranchuk, Dmitry and Rubachev, Ivan and Babenko, Artem},
  booktitle={International Conference on Machine Learning},
  pages={17564--17579},
  year={2023},
  organization={PMLR}
}

@article{synth_2,
  title={Challenges and opportunities of generative models on tabular data},
  author={Wang, Alex X. and Chukova, Stefanka S. and Simpson, Colin R. and Nguyen, Binh P.},
  journal={Applied Soft Computing},
  pages={112223},
  year={2024},
  publisher={Elsevier}
}

@article{tabpfn,
  title={{TabPFN}: a transformer that solves small tabular classification problems in a second},
  author={Hollmann, Noah and M{\"u}ller, Samuel and Eggensperger, Katharina and Hutter, Frank},
  journal={arXiv preprint arXiv:2207.01848},
  year={2022}
}

@article{guo2022,
  title={Multiparametric {MRI}-based radiomics model for predicting {H3 K27M} mutant status in diffuse midline glioma: a comparative study across different sequences and machine learning techniques},
  author={Guo, Weina and She, Dejun and Xing, Zhen and Lin, Xiefeng and Wang, Fei and Song, Yang and Cao, Dairong},
  journal={Frontiers in Oncology},
  volume={12},
  pages={796583},
  year={2022},
  publisher={Frontiers Media SA}
}

@article{indoria2024,
  title={Prediction of {H3K27M} alteration in midline gliomas of the brain using radiomics: a multi-institute study},
  author={Indoria, Akshay and Arora, Aditi and Garg, Ajay and Chauhan, Richa Singh and Chaturvedi, Arvinda and Kumar, Maya and Konar, Subhas and Sadashiva, Nishanth and Rao, Shilpa and Saini, Jitender},
  journal={Neuro-Oncology Advances},
  volume={6},
  number={1},
  pages={vdae153},
  year={2024},
  publisher={Oxford University Press}
}

@inproceedings{slicer,
  title={{3D Slicer}},
  author={Pieper, Steve and Halle, Michael and Kikinis, Ron},
  booktitle={2004 2nd IEEE International Symposium on Biomedical Imaging: Nano to Macro},
  pages={632--635},
  year={2004},
  organization={IEEE}
}

@inproceedings{bet2,
  title={{BET2}: {MR}-based estimation of brain, skull and scalp surfaces},
  author={Jenkinson, Mark},
  booktitle={Eleventh Annual Meeting of the Organization for Human Brain Mapping},
  year={2005}
}

@article{pyRadiomics,
  title={Computational radiomics system to decode the radiographic phenotype},
  author={Van Griethuysen, Joost J. M. and Fedorov, Andriy and Parmar, Chintan and Hosny, Ahmed and Aucoin, Nicole and Narayan, Vivek and Beets-Tan, Regina G. H. and Fillion-Robin, Jean-Christophe and Pieper, Steve and Aerts, Hugo J. W. L.},
  journal={Cancer Research},
  volume={77},
  number={21},
  pages={e104--e107},
  year={2017},
  publisher={AACR}
}

@article{smote,
  title={{SMOTE}: synthetic minority over-sampling technique},
  author={Chawla, Nitesh V. and Bowyer, Kevin W. and Hall, Lawrence O. and Kegelmeyer, W. Philip},
  journal={Journal of Artificial Intelligence Research},
  volume={16},
  pages={321--357},
  year={2002}
}

@article{tripodai,
  title={{TRIPOD+AI} statement: updated guidance for reporting clinical prediction models that use regression or machine learning methods},
  author={Collins, Gary S. and Moons, Karel G. M. and Dhiman, Paula and Riley, Richard D. and Beam, Andrew L. and Van Calster, Ben and Ghassemi, Marzyeh and Liu, Xiaoxuan and Reitsma, Johannes B. and Van Smeden, Maarten and Boulesteix, Anne-Laure and Camaradou, Jennifer Catherine and Celi, Leo Anthony and Denaxas, Spiros and Denniston, Alastair K. and Glocker, Ben and Golub, Robert M. and Harvey, Hugh and Heinze, Georg and Hoffman, Michael M. and Kengne, Andr{\'e} Pascal and Lam, Emily and Lee, Naomi and Loder, Elizabeth W. and Maier-Hein, Lena and Mateen, Bilal A. and McCradden, Melissa D. and Oakden-Rayner, Lauren and Ordish, Johan and Parnell, Richard and Rose, Sherri and Singh, Karandeep and Wynants, Laure and Logullo, Patricia},
  journal={BMJ},
  volume={385},
  pages={e078378},
  year={2024},
  doi={10.1136/bmj-2023-078378}
}

\section*{Acknowledgements}
The authors thank the Russian Research Center of Roentgenology and Radiology for providing anonymized imaging data, molecular labels, and expert tumor segmentations for this retrospective study.

\section*{Author contributions statement}
A.Z. developed the MRI preprocessing and standardization workflow, feature-space expansion, feature-selection pipeline, and TabDDPM synthetic radiomic augmentation pipeline, and contributed to the literature review and manuscript drafting. N.K. and T.K. coordinated data acquisition with clinical collaborators, performed radiomic feature extraction and data standardization, integrated tumor-volume data, contributed to skull-stripping and unlabeled-data experiments, and contributed to clinical interpretation and manuscript drafting. S.P. implemented the repeated-split evaluation framework, trained and evaluated the machine-learning classifiers, ran the main experimental pipeline, and prepared the results tables and figures. V.B.  made expert tumor segmentations for this retrospective study. A.B. and Y.K. supervised the study, contributed to study design and methodological interpretation, and reviewed the manuscript. All authors reviewed and approved the final manuscript. 

\section*{Additional information}
\textbf{Competing interests} The author(s) declare no competing interests.

\end{document}


\maketitle

\section*{Supplementary Methods}

\subsection*{Radiomics extraction settings}
PyRadiomics was used with default settings \cite{pyRadiomics}. Features were extracted from the tumor region of interest in 3D and included shape, first-order statistics, and texture features derived from the gray-level co-occurrence matrix (GLCM), gray-level run length matrix (GLRLM), gray-level size zone matrix (GLSZM), and gray-level dependence matrix (GLDM). In the preprocessing pipeline, features were extracted from the original image and from filtered images, including wavelet sub-bands, Laplacian-of-Gaussian images with $\sigma = 1.0$ and $3.0$, and nonlinear intensity transforms.

\subsection*{TabDDPM training and evaluation details}
TabDDPM was trained on the 10-feature subset selected by correlation pruning and repeated recursive feature elimination (RFE). Conditional generation was used with two classes (H3K27M-positive and wild-type), and synthetic samples were added to the training folds only to balance classes. Hyperparameters for the diffusion model and its CatBoost evaluator are listed in Supplementary Tables~\ref{tab:S14_catboost_hparams} and \ref{tab:S15_ddpm_hparams}.

\section*{Supplementary Results}
Unless otherwise stated, all performance values in the supplementary tables are reported as mean $\pm$ standard deviation across 100 random seeds using stratified 70/30 train/test splits. Test sets always consisted of real (non-synthetic) cases only. In all performance tables, bold text indicates the best value, and underlined text indicates the second-best value within each metric column (by mean).

\subsection*{Experiment 1. Pure radiomics and volumetric augmentation}
This experiment evaluated the baseline predictive value of radiomic features extracted from the original T2-weighted images, before applying preprocessing, feature selection, or synthetic augmentation. The first table reports the pure-radiomics baseline; the second table shows the effect of adding tumor volume as an explicit scalar descriptor.

\begin{table}[H]
\centering
\caption{Experiment 1A. Pure radiomics baseline. Performance of each classifier using radiomic features extracted from the original T2-weighted images only, without preprocessing, feature selection, or synthetic augmentation. Values are mean $\pm$ SD across 100 random splits.}
\label{tab:S1}
\scriptsize
\begin{tabularx}{\textwidth}{Xcccc}
\toprule
\textbf{Model} & \textbf{Accuracy} & \textbf{Recall} & \textbf{Precision} & \textbf{F1-score} \\
\midrule
TabPFN & 0.66 $\pm$ 0.08 & 0.83 $\pm$ 0.10 & 0.75 $\pm$ 0.05 & 0.78 $\pm$ 0.06 \\
Adaptive Boosting & 0.63 $\pm$ 0.09 & 0.72 $\pm$ 0.11 & 0.77 $\pm$ 0.06 & 0.74 $\pm$ 0.07 \\
Logistic Regression & 0.60 $\pm$ 0.13 & 0.71 $\pm$ 0.21 & 0.74 $\pm$ 0.07 & 0.71 $\pm$ 0.15 \\
LightGBM & 0.66 $\pm$ 0.09 & 0.76 $\pm$ 0.11 & 0.79 $\pm$ 0.06 & 0.77 $\pm$ 0.07 \\
K-nearest neighbors & 0.58 $\pm$ 0.10 & 0.58 $\pm$ 0.13 & 0.81 $\pm$ 0.09 & 0.67 $\pm$ 0.10 \\
SVM & 0.66 $\pm$ 0.20 & 0.81 $\pm$ 0.36 & 0.69 $\pm$ 0.21 & 0.72 $\pm$ 0.29 \\
Random Forest Classifier & 0.59 $\pm$ 0.09 & 0.70 $\pm$ 0.10 & 0.74 $\pm$ 0.06 & 0.72 $\pm$ 0.07 \\
Decision Tree & 0.63 $\pm$ 0.11 & 0.70 $\pm$ 0.13 & 0.79 $\pm$ 0.08 & 0.74 $\pm$ 0.09 \\
Ridge Logistic Regression & 0.53 $\pm$ 0.09 & 0.58 $\pm$ 0.12 & 0.74 $\pm$ 0.07 & 0.64 $\pm$ 0.09 \\
AutoML & 0.57 $\pm$ 0.09 & 0.62 $\pm$ 0.10 & 0.77 $\pm$ 0.11 & 0.68 $\pm$ 0.09 \\
CatBoost & 0.62 $\pm$ 0.09 & 0.74 $\pm$ 0.12 & 0.75 $\pm$ 0.06 & 0.74 $\pm$ 0.07 \\
XGBoost & 0.66 $\pm$ 0.09 & 0.76 $\pm$ 0.11 & 0.78 $\pm$ 0.06 & 0.77 $\pm$ 0.07 \\
\bottomrule
\end{tabularx}
\end{table}

\begin{table}[H]
\centering
\caption{Experiment 1B. Pure radiomics plus tumor volume. Performance after appending tumor volume to the original radiomics feature set, without preprocessing, feature selection, or synthetic augmentation. Values are mean $\pm$ SD across 100 random splits.}
\label{tab:S2}
\scriptsize
\begin{tabularx}{\textwidth}{Xcccc}
\toprule
\textbf{Model} & \textbf{Accuracy} & \textbf{Recall} & \textbf{Precision} & \textbf{F1-score} \\
\midrule
TabPFN & 0.67 $\pm$ 0.09 & 0.85 $\pm$ 0.11 & 0.72 $\pm$ 0.06 & 0.78 $\pm$ 0.07 \\
Adaptive Boosting & 0.64 $\pm$ 0.10 & 0.75 $\pm$ 0.13 & 0.75 $\pm$ 0.07 & 0.74 $\pm$ 0.08 \\
Logistic Regression & 0.58 $\pm$ 0.12 & 0.69 $\pm$ 0.23 & 0.71 $\pm$ 0.07 & 0.68 $\pm$ 0.16 \\
LightGBM & 0.68 $\pm$ 0.10 & 0.78 $\pm$ 0.13 & 0.77 $\pm$ 0.07 & 0.77 $\pm$ 0.08 \\
K-nearest neighbors & 0.61 $\pm$ 0.11 & 0.64 $\pm$ 0.14 & 0.78 $\pm$ 0.10 & 0.69 $\pm$ 0.10 \\
SVM & 0.64 $\pm$ 0.17 & 0.81 $\pm$ 0.36 & 0.65 $\pm$ 0.22 & 0.69 $\pm$ 0.29 \\
Random Forest Classifier & 0.63 $\pm$ 0.10 & 0.76 $\pm$ 0.13 & 0.73 $\pm$ 0.07 & 0.74 $\pm$ 0.09 \\
Decision Tree & 0.62 $\pm$ 0.11 & 0.71 $\pm$ 0.13 & 0.74 $\pm$ 0.08 & 0.72 $\pm$ 0.09 \\
Ridge Logistic Regression & 0.59 $\pm$ 0.13 & 0.64 $\pm$ 0.21 & 0.74 $\pm$ 0.09 & 0.67 $\pm$ 0.15 \\
AutoML & 0.57 $\pm$ 0.09 & 0.62 $\pm$ 0.10 & 0.77 $\pm$ 0.11 & 0.68 $\pm$ 0.09 \\
CatBoost & 0.65 $\pm$ 0.09 & 0.78 $\pm$ 0.12 & 0.74 $\pm$ 0.06 & 0.76 $\pm$ 0.08 \\
XGBoost & 0.67 $\pm$ 0.10 & 0.78 $\pm$ 0.13 & 0.76 $\pm$ 0.07 & 0.76 $\pm$ 0.08 \\
\bottomrule
\end{tabularx}
\end{table}

\subsection*{Experiment 2. Effects of preprocessing}
This experiment examined whether preprocessing improved robustness and predictive performance in the heterogeneous T2-weighted dataset. Preprocessing included isotropic resampling, skull stripping, and filtered-image feature extraction. Supplementary Tables~\ref{tab:S3} and \ref{tab:S4} compare preprocessing without and with the explicit tumor volume feature.

\begin{table}[H]
\centering
\caption{Experiment 2A. Preprocessing only. Performance after applying the preprocessing pipeline, without explicit tumor volume, feature selection, or synthetic augmentation. Values are mean $\pm$ SD across 100 random splits.}
\label{tab:S3}
\scriptsize
\begin{tabularx}{\textwidth}{Xcccc}
\toprule
\textbf{Model} & \textbf{Accuracy} & \textbf{Recall} & \textbf{Precision} & \textbf{F1-score} \\
\midrule
TabPFN & 0.68 $\pm$ 0.08 & 0.83 $\pm$ 0.10 & 0.76 $\pm$ 0.04 & 0.79 $\pm$ 0.06 \\
Adaptive Boosting & 0.66 $\pm$ 0.09 & 0.78 $\pm$ 0.11 & 0.76 $\pm$ 0.06 & 0.77 $\pm$ 0.07 \\
Logistic Regression & 0.67 $\pm$ 0.10 & 0.72 $\pm$ 0.12 & 0.82 $\pm$ 0.08 & 0.76 $\pm$ 0.09 \\
LightGBM & 0.67 $\pm$ 0.09 & 0.81 $\pm$ 0.11 & 0.77 $\pm$ 0.05 & 0.79 $\pm$ 0.07 \\
K-nearest neighbors & 0.62 $\pm$ 0.10 & 0.65 $\pm$ 0.13 & 0.81 $\pm$ 0.08 & 0.71 $\pm$ 0.10 \\
SVM & 0.61 $\pm$ 0.22 & 0.71 $\pm$ 0.41 & 0.71 $\pm$ 0.20 & 0.64 $\pm$ 0.32 \\
Random Forest Classifier & 0.68 $\pm$ 0.08 & 0.81 $\pm$ 0.11 & 0.78 $\pm$ 0.05 & 0.79 $\pm$ 0.07 \\
Decision Tree & 0.62 $\pm$ 0.10 & 0.71 $\pm$ 0.13 & 0.76 $\pm$ 0.07 & 0.73 $\pm$ 0.09 \\
Ridge Logistic Regression & 0.58 $\pm$ 0.10 & 0.62 $\pm$ 0.12 & 0.78 $\pm$ 0.08 & 0.69 $\pm$ 0.09 \\
AutoML & 0.65 $\pm$ 0.10 & 0.80 $\pm$ 0.10 & 0.78 $\pm$ 0.09 & 0.78 $\pm$ 0.07 \\
CatBoost & 0.69 $\pm$ 0.08 & 0.82 $\pm$ 0.10 & 0.78 $\pm$ 0.05 & 0.80 $\pm$ 0.06 \\
XGBoost & 0.68 $\pm$ 0.08 & 0.81 $\pm$ 0.11 & 0.77 $\pm$ 0.05 & 0.79 $\pm$ 0.07 \\
\bottomrule
\end{tabularx}
\end{table}

\begin{table}[H]
\centering
\caption{Experiment 2B. Preprocessing plus tumor volume. Performance after preprocessing and inclusion of tumor volume, without feature selection or synthetic augmentation. Values are mean $\pm$ SD across 100 random splits.}
\label{tab:S4}
\scriptsize
\begin{tabularx}{\textwidth}{Xcccc}
\toprule
\textbf{Model} & \textbf{Accuracy} & \textbf{Recall} & \textbf{Precision} & \textbf{F1-score} \\
\midrule
TabPFN & 0.67 $\pm$ 0.08 & 0.85 $\pm$ 0.10 & 0.73 $\pm$ 0.06 & 0.78 $\pm$ 0.06 \\
Adaptive Boosting & 0.62 $\pm$ 0.10 & 0.75 $\pm$ 0.12 & 0.72 $\pm$ 0.07 & 0.73 $\pm$ 0.08 \\
Logistic Regression & 0.65 $\pm$ 0.10 & 0.74 $\pm$ 0.13 & 0.76 $\pm$ 0.08 & 0.75 $\pm$ 0.09 \\
LightGBM & 0.64 $\pm$ 0.09 & 0.78 $\pm$ 0.11 & 0.73 $\pm$ 0.07 & 0.75 $\pm$ 0.07 \\
K-nearest neighbors & 0.63 $\pm$ 0.11 & 0.67 $\pm$ 0.17 & 0.77 $\pm$ 0.09 & 0.71 $\pm$ 0.12 \\
SVM & 0.59 $\pm$ 0.19 & 0.72 $\pm$ 0.38 & 0.66 $\pm$ 0.20 & 0.64 $\pm$ 0.29 \\
Random Forest Classifier & 0.65 $\pm$ 0.10 & 0.78 $\pm$ 0.13 & 0.74 $\pm$ 0.07 & 0.75 $\pm$ 0.08 \\
Decision Tree & 0.58 $\pm$ 0.10 & 0.68 $\pm$ 0.13 & 0.71 $\pm$ 0.08 & 0.69 $\pm$ 0.09 \\
Ridge Logistic Regression & 0.52 $\pm$ 0.12 & 0.55 $\pm$ 0.15 & 0.70 $\pm$ 0.11 & 0.60 $\pm$ 0.12 \\
AutoML & 0.65 $\pm$ 0.09 & 0.77 $\pm$ 0.09 & 0.72 $\pm$ 0.08 & 0.75 $\pm$ 0.06 \\
CatBoost & 0.65 $\pm$ 0.10 & 0.78 $\pm$ 0.13 & 0.74 $\pm$ 0.06 & 0.75 $\pm$ 0.08 \\
XGBoost & 0.61 $\pm$ 0.09 & 0.76 $\pm$ 0.12 & 0.71 $\pm$ 0.06 & 0.73 $\pm$ 0.07 \\
\bottomrule
\end{tabularx}
\end{table}

\subsection*{Experiment 3. Feature selection}
After preprocessing and filtering, feature dimensionality increased substantially. This experiment assessed whether correlation pruning and repeated RFE improved stability and performance. Supplementary Table~\ref{tab:S5_feature_ranking} lists the 10 most stable selected features. Supplementary Tables~\ref{tab:S6} and \ref{tab:S7} report downstream classifier performance without and with tumor volume.

\begin{table}[H]
\centering
\caption{Experiment 3A. Top 10 selected features. Mean ranking statistics and frequency of appearance in the top 10 across 100 RFE runs.}
\label{tab:S5_feature_ranking}
\scriptsize
\begin{tabularx}{\textwidth}{Xcc}
\toprule
\textbf{Feature} & \textbf{Mean Rank} & \textbf{Freq. in Top 10} \\
\midrule
logarithm\_gldm\_SmallDependenceLowGrayLevelEmphasis\_only\_brain & 7.3 $\pm$ 9.0 & 0.8 \\
square\_gldm\_DependenceEntropy\_only\_brain & 10.5 $\pm$ 11.9 & 0.7 \\
square\_glszm\_SmallAreaLowGrayLevelEmphasis\_only\_brain & 14.2 $\pm$ 32.7 & 0.7 \\
original\_shape\_SurfaceVolumeRatio\_skull\_stripped & 18.0 $\pm$ 19.8 & 0.4 \\
exponential\_glcm\_InverseVariance\_only\_brain & 18.8 $\pm$ 40.0 & 0.7 \\
wavelet-HLL\_glcm\_Correlation\_only\_brain & 19.4 $\pm$ 25.8 & 0.5 \\
exponential\_glszm\_SmallAreaLowGrayLevelEmphasis\_skull\_stripped & 19.9 $\pm$ 29.6 & 0.5 \\
wavelet-HLH\_glcm\_Idn\_skull\_stripped & 20.5 $\pm$ 24.5 & 0.4 \\
exponential\_glcm\_Idn\_only\_brain & 22.6 $\pm$ 39.1 & 0.5 \\
wavelet-HLH\_glcm\_Idmn\_only\_brain & 22.9 $\pm$ 30.9 & 0.4 \\

\bottomrule
\end{tabularx}
\end{table}

\begin{table}[H]
\centering
\caption{Experiment 3B. Feature selection without tumor volume. Performance after correlation pruning and repeated RFE, without synthetic augmentation. Values are mean $\pm$ SD across 100 random splits.}
\label{tab:S6}
\scriptsize
\begin{tabularx}{\textwidth}{Xcccc}
\toprule
\textbf{Model} & \textbf{Accuracy} & \textbf{Recall} & \textbf{Precision} & \textbf{F1-score} \\
\midrule
TabPFN & 0.68 $\pm$ 0.09 & 0.86 $\pm$ 0.11 & 0.74 $\pm$ 0.06 & 0.79 $\pm$ 0.07 \\
Adaptive Boosting & 0.68 $\pm$ 0.11 & 0.78 $\pm$ 0.14 & 0.77 $\pm$ 0.07 & 0.77 $\pm$ 0.09 \\
Logistic Regression & 0.63 $\pm$ 0.10 & 0.67 $\pm$ 0.12 & 0.77 $\pm$ 0.08 & 0.71 $\pm$ 0.09 \\
LightGBM & 0.66 $\pm$ 0.10 & 0.77 $\pm$ 0.13 & 0.76 $\pm$ 0.08 & 0.76 $\pm$ 0.09 \\
K-nearest neighbors & 0.61 $\pm$ 0.10 & 0.68 $\pm$ 0.14 & 0.75 $\pm$ 0.06 & 0.70 $\pm$ 0.10 \\
SVM & 0.47 $\pm$ 0.11 & 0.38 $\pm$ 0.19 & 0.72 $\pm$ 0.23 & 0.47 $\pm$ 0.19 \\
Random Forest Classifier & 0.70 $\pm$ 0.09 & 0.82 $\pm$ 0.12 & 0.77 $\pm$ 0.07 & 0.79 $\pm$ 0.08 \\
Decision Tree & 0.63 $\pm$ 0.10 & 0.72 $\pm$ 0.14 & 0.75 $\pm$ 0.08 & 0.73 $\pm$ 0.09 \\
Ridge Logistic Regression & 0.60 $\pm$ 0.09 & 0.69 $\pm$ 0.12 & 0.74 $\pm$ 0.07 & 0.71 $\pm$ 0.08 \\
AutoML & 0.63 $\pm$ 0.19 & 0.75 $\pm$ 0.40 & 0.59 $\pm$ 0.30 & 0.66 $\pm$ 0.34 \\
CatBoost & 0.68 $\pm$ 0.09 & 0.81 $\pm$ 0.12 & 0.76 $\pm$ 0.07 & 0.78 $\pm$ 0.07 \\
XGBoost & 0.67 $\pm$ 0.10 & 0.79 $\pm$ 0.12 & 0.76 $\pm$ 0.07 & 0.77 $\pm$ 0.08 \\
\bottomrule
\end{tabularx}
\end{table}

\begin{table}[H]
\centering
\caption{Experiment 3C. Feature selection plus tumor volume. Performance after correlation pruning, repeated RFE, and inclusion of tumor volume, without synthetic augmentation. Values are mean $\pm$ SD across 100 random splits.}
\label{tab:S7}
\scriptsize
\begin{tabularx}{\textwidth}{Xcccc}
\toprule
\textbf{Model} & \textbf{Accuracy} & \textbf{Recall} & \textbf{Precision} & \textbf{F1-score} \\
\midrule
TabPFN & 0.69 $\pm$ 0.07 & 0.88 $\pm$ 0.07 & 0.74 $\pm$ 0.09 & 0.80 $\pm$ 0.06 \\
Adaptive Boosting & 0.69 $\pm$ 0.11 & 0.81 $\pm$ 0.10 & 0.79 $\pm$ 0.06 & 0.79 $\pm$ 0.08 \\
Logistic Regression & 0.62 $\pm$ 0.08 & 0.65 $\pm$ 0.08 & 0.81 $\pm$ 0.06 & 0.72 $\pm$ 0.07 \\
LightGBM & 0.68 $\pm$ 0.08 & 0.79 $\pm$ 0.09 & 0.78 $\pm$ 0.04 & 0.78 $\pm$ 0.06 \\
K-nearest neighbors & 0.63 $\pm$ 0.08 & 0.69 $\pm$ 0.10 & 0.80 $\pm$ 0.07 & 0.73 $\pm$ 0.07 \\
SVM & 0.67 $\pm$ 0.08 & 0.72 $\pm$ 0.12 & 0.83 $\pm$ 0.08 & 0.76 $\pm$ 0.07 \\
Random Forest Classifier & 0.71 $\pm$ 0.08 & 0.84 $\pm$ 0.08 & 0.79 $\pm$ 0.05 & 0.81 $\pm$ 0.06 \\
Decision Tree & 0.65 $\pm$ 0.07 & 0.78 $\pm$ 0.07 & 0.76 $\pm$ 0.04 & 0.77 $\pm$ 0.05 \\
Ridge Logistic Regression & 0.65 $\pm$ 0.10 & 0.72 $\pm$ 0.12 & 0.79 $\pm$ 0.05 & 0.75 $\pm$ 0.09 \\
AutoML & 0.70 $\pm$ 0.13 & 0.83 $\pm$ 0.09 & 0.79 $\pm$ 0.14 & 0.80 $\pm$ 0.09 \\
CatBoost & 0.73 $\pm$ 0.07 & 0.85 $\pm$ 0.06 & 0.80 $\pm$ 0.05 & 0.83 $\pm$ 0.04 \\
XGBoost & 0.69 $\pm$ 0.11 & 0.81 $\pm$ 0.09 & 0.79 $\pm$ 0.07 & 0.80 $\pm$ 0.08 \\
\bottomrule
\end{tabularx}
\end{table}

\subsection*{Experiment 4. Synthetic augmentation with TabDDPM}
This experiment assessed whether diffusion-based tabular augmentation improved classification after preprocessing, feature selection, and inclusion of tumor volume. Among the evaluated classifiers, the clearest benefit was observed for TabPFN, whose best synthetic configuration occurred at 200 augmented rows. The full per-model performance values for all augmentation sizes are reported in Supplementary Tables~\ref{tab:S8}--\ref{tab:S13}.

\begin{table}[H]
\centering
\caption{Experiment 4A. Synthetic augmentation to 94 total rows after feature selection and volume inclusion. Values are mean $\pm$ SD across 100 random splits.}
\label{tab:S8}
\scriptsize
\begin{tabularx}{\textwidth}{Xcccc}
\toprule
\textbf{Model} & \textbf{Accuracy} & \textbf{Recall} & \textbf{Precision} & \textbf{F1-score} \\
\midrule
TabPFN & 0.69 $\pm$ 0.06 & 0.79 $\pm$ 0.01 & 0.72 $\pm$ 0.07 & 0.75 $\pm$ 0.04 \\
Adaptive Boosting & 0.59 $\pm$ 0.09 & 0.72 $\pm$ 0.17 & 0.61 $\pm$ 0.06 & 0.66 $\pm$ 0.10 \\
Logistic Regression & 0.53 $\pm$ 0.09 & 0.89 $\pm$ 0.11 & 0.55 $\pm$ 0.05 & 0.68 $\pm$ 0.07 \\
LightGBM & 0.56 $\pm$ 0.12 & 0.72 $\pm$ 0.17 & 0.58 $\pm$ 0.08 & 0.64 $\pm$ 0.12 \\
K-nearest neighbors & 0.59 $\pm$ 0.03 & 0.94 $\pm$ 0.06 & 0.59 $\pm$ 0.01 & 0.72 $\pm$ 0.03 \\
SVM & 0.56 $\pm$ 0.00 & 0.94 $\pm$ 0.01 & 0.56 $\pm$ 0.00 & 0.70 $\pm$ 0.00 \\
Random Forest Classifier & 0.61 $\pm$ 0.03 & 0.85 $\pm$ 0.07 & 0.67 $\pm$ 0.04 & 0.74 $\pm$ 0.03 \\
Decision Tree & 0.57 $\pm$ 0.07 & 0.71 $\pm$ 0.05 & 0.60 $\pm$ 0.06 & 0.65 $\pm$ 0.06 \\
Ridge Logistic Regression & 0.47 $\pm$ 0.03 & 0.83 $\pm$ 0.06 & 0.52 $\pm$ 0.02 & 0.64 $\pm$ 0.03 \\
CatBoost & 0.56 $\pm$ 0.00 & 0.89 $\pm$ 0.00 & 0.57 $\pm$ 0.00 & 0.70 $\pm$ 0.00 \\
XGBoost & 0.66 $\pm$ 0.03 & 0.89 $\pm$ 0.00 & 0.64 $\pm$ 0.03 & 0.74 $\pm$ 0.02 \\
\bottomrule
\end{tabularx}
\end{table}

\begin{table}[H]
\centering
\caption{Experiment 4B. Synthetic augmentation to 200 total rows after feature selection and volume inclusion. Values are mean $\pm$ SD across 100 random splits.}
\label{tab:S9}
\scriptsize
\begin{tabularx}{\textwidth}{Xcccc}
\toprule
\textbf{Model} & \textbf{Accuracy} & \textbf{Recall} & \textbf{Precision} & \textbf{F1-score} \\
\midrule
TabPFN & 0.71 $\pm$ 0.10 & 0.85 $\pm$ 0.05 & 0.78 $\pm$ 0.09 & 0.81 $\pm$ 0.05 \\
Adaptive Boosting & 0.69 $\pm$ 0.13 & 0.81 $\pm$ 0.12 & 0.76 $\pm$ 0.14 & 0.78 $\pm$ 0.11 \\
Logistic Regression & 0.63 $\pm$ 0.12 & 0.80 $\pm$ 0.16 & 0.69 $\pm$ 0.10 & 0.73 $\pm$ 0.12 \\
LightGBM & 0.69 $\pm$ 0.10 & 0.79 $\pm$ 0.12 & 0.76 $\pm$ 0.11 & 0.77 $\pm$ 0.09 \\
K-nearest neighbors & 0.65 $\pm$ 0.09 & 0.83 $\pm$ 0.12 & 0.71 $\pm$ 0.09 & 0.76 $\pm$ 0.08 \\
SVM & 0.68 $\pm$ 0.08 & 0.99 $\pm$ 0.03 & 0.68 $\pm$ 0.08 & 0.80 $\pm$ 0.06 \\
Random Forest Classifier & 0.65 $\pm$ 0.10 & 0.79 $\pm$ 0.07 & 0.74 $\pm$ 0.11 & 0.76 $\pm$ 0.06 \\
Decision Tree & 0.63 $\pm$ 0.11 & 0.73 $\pm$ 0.13 & 0.72 $\pm$ 0.13 & 0.72 $\pm$ 0.12 \\
Ridge Logistic Regression & 0.60 $\pm$ 0.13 & 0.76 $\pm$ 0.16 & 0.67 $\pm$ 0.12 & 0.71 $\pm$ 0.12 \\
CatBoost & 0.63 $\pm$ 0.12 & 0.78 $\pm$ 0.07 & 0.71 $\pm$ 0.13 & 0.74 $\pm$ 0.09 \\
XGBoost & 0.64 $\pm$ 0.09 & 0.77 $\pm$ 0.09 & 0.73 $\pm$ 0.11 & 0.74 $\pm$ 0.08 \\
\bottomrule
\end{tabularx}
\end{table}

\begin{table}[H]
\centering
\caption{Experiment 4C. Synthetic augmentation to 400 total rows after feature selection and volume inclusion. Values are mean $\pm$ SD across 100 random splits.}
\label{tab:S10}
\scriptsize
\begin{tabularx}{\textwidth}{Xcccc}
\toprule
\textbf{Model} & \textbf{Accuracy} & \textbf{Recall} & \textbf{Precision} & \textbf{F1-score} \\
\midrule
TabPFN & 0.71 $\pm$ 0.10 & 0.81 $\pm$ 0.05 & 0.77 $\pm$ 0.11 & 0.79 $\pm$ 0.07 \\
Adaptive Boosting & 0.62 $\pm$ 0.10 & 0.72 $\pm$ 0.08 & 0.73 $\pm$ 0.13 & 0.72 $\pm$ 0.08 \\
Logistic Regression & 0.62 $\pm$ 0.13 & 0.74 $\pm$ 0.15 & 0.70 $\pm$ 0.11 & 0.71 $\pm$ 0.12 \\
LightGBM & 0.67 $\pm$ 0.13 & 0.76 $\pm$ 0.09 & 0.76 $\pm$ 0.13 & 0.75 $\pm$ 0.10 \\
K-nearest neighbors & 0.59 $\pm$ 0.09 & 0.74 $\pm$ 0.15 & 0.69 $\pm$ 0.06 & 0.70 $\pm$ 0.07 \\
SVM & 0.68 $\pm$ 0.09 & 0.96 $\pm$ 0.07 & 0.69 $\pm$ 0.08 & 0.80 $\pm$ 0.07 \\
Random Forest Classifier & 0.66 $\pm$ 0.10 & 0.80 $\pm$ 0.10 & 0.73 $\pm$ 0.10 & 0.76 $\pm$ 0.09 \\
Decision Tree & 0.66 $\pm$ 0.10 & 0.73 $\pm$ 0.10 & 0.76 $\pm$ 0.13 & 0.74 $\pm$ 0.10 \\
Ridge Logistic Regression & 0.64 $\pm$ 0.12 & 0.73 $\pm$ 0.14 & 0.73 $\pm$ 0.10 & 0.72 $\pm$ 0.11 \\
CatBoost & 0.66 $\pm$ 0.10 & 0.78 $\pm$ 0.10 & 0.73 $\pm$ 0.12 & 0.75 $\pm$ 0.09 \\
XGBoost & 0.66 $\pm$ 0.11 & 0.77 $\pm$ 0.09 & 0.74 $\pm$ 0.12 & 0.75 $\pm$ 0.09 \\
AutoML & 0.69 $\pm$ 0.00 & 0.78 $\pm$ 0.00 & 0.70 $\pm$ 0.00 & 0.74 $\pm$ 0.00 \\
\bottomrule
\end{tabularx}
\end{table}

\begin{table}[H]
\centering
\caption{Experiment 4D. Synthetic augmentation to 600 total rows after feature selection and volume inclusion. Values are mean $\pm$ SD across 100 random splits.}
\label{tab:S11}
\scriptsize
\begin{tabularx}{\textwidth}{Xcccc}
\toprule
\textbf{Model} & \textbf{Accuracy} & \textbf{Recall} & \textbf{Precision} & \textbf{F1-score} \\
\midrule
TabPFN & 0.70 $\pm$ 0.10 & 0.80 $\pm$ 0.06 & 0.77 $\pm$ 0.11 & 0.78 $\pm$ 0.08 \\
Adaptive Boosting & 0.63 $\pm$ 0.12 & 0.74 $\pm$ 0.16 & 0.72 $\pm$ 0.13 & 0.72 $\pm$ 0.13 \\
Logistic Regression & 0.63 $\pm$ 0.12 & 0.72 $\pm$ 0.15 & 0.72 $\pm$ 0.11 & 0.71 $\pm$ 0.12 \\
LightGBM & 0.66 $\pm$ 0.13 & 0.76 $\pm$ 0.09 & 0.76 $\pm$ 0.14 & 0.75 $\pm$ 0.10 \\
K-nearest neighbors & 0.59 $\pm$ 0.11 & 0.74 $\pm$ 0.16 & 0.68 $\pm$ 0.08 & 0.70 $\pm$ 0.10 \\
SVM & 0.67 $\pm$ 0.12 & 0.84 $\pm$ 0.29 & 0.64 $\pm$ 0.23 & 0.73 $\pm$ 0.25 \\
Random Forest Classifier & 0.66 $\pm$ 0.10 & 0.79 $\pm$ 0.10 & 0.73 $\pm$ 0.10 & 0.75 $\pm$ 0.09 \\
Decision Tree & 0.61 $\pm$ 0.14 & 0.70 $\pm$ 0.12 & 0.72 $\pm$ 0.15 & 0.70 $\pm$ 0.12 \\
Ridge Logistic Regression & 0.64 $\pm$ 0.12 & 0.73 $\pm$ 0.14 & 0.74 $\pm$ 0.10 & 0.73 $\pm$ 0.10 \\
AutoML & 0.69 $\pm$ 0.00 & 0.78 $\pm$ 0.00 & 0.70 $\pm$ 0.00 & 0.74 $\pm$ 0.00 \\
CatBoost & 0.66 $\pm$ 0.11 & 0.79 $\pm$ 0.10 & 0.73 $\pm$ 0.11 & 0.75 $\pm$ 0.10 \\
XGBoost & 0.66 $\pm$ 0.10 & 0.77 $\pm$ 0.11 & 0.74 $\pm$ 0.11 & 0.75 $\pm$ 0.08 \\
\bottomrule
\end{tabularx}
\end{table}

\begin{table}[H]
\centering
\caption{Experiment 4E. Synthetic augmentation to 800 total rows after feature selection and volume inclusion. Values are mean $\pm$ SD across 100 random splits.}
\label{tab:S12}
\scriptsize
\begin{tabularx}{\textwidth}{Xcccc}
\toprule
\textbf{Model} & \textbf{Accuracy} & \textbf{Recall} & \textbf{Precision} & \textbf{F1-score} \\
\midrule
TabPFN & 0.69 $\pm$ 0.11 & 0.77 $\pm$ 0.09 & 0.77 $\pm$ 0.12 & 0.77 $\pm$ 0.09 \\
Adaptive Boosting & 0.64 $\pm$ 0.08 & 0.74 $\pm$ 0.08 & 0.74 $\pm$ 0.10 & 0.73 $\pm$ 0.07 \\
Logistic Regression & 0.64 $\pm$ 0.12 & 0.73 $\pm$ 0.14 & 0.73 $\pm$ 0.11 & 0.72 $\pm$ 0.11 \\
LightGBM & 0.67 $\pm$ 0.13 & 0.76 $\pm$ 0.13 & 0.75 $\pm$ 0.14 & 0.75 $\pm$ 0.12 \\
K-nearest neighbors & 0.57 $\pm$ 0.10 & 0.73 $\pm$ 0.15 & 0.67 $\pm$ 0.08 & 0.69 $\pm$ 0.10 \\
SVM & 0.64 $\pm$ 0.12 & 0.79 $\pm$ 0.28 & 0.63 $\pm$ 0.23 & 0.70 $\pm$ 0.25 \\
Random Forest Classifier & 0.67 $\pm$ 0.09 & 0.79 $\pm$ 0.09 & 0.73 $\pm$ 0.10 & 0.76 $\pm$ 0.08 \\
Decision Tree & 0.65 $\pm$ 0.15 & 0.70 $\pm$ 0.15 & 0.75 $\pm$ 0.15 & 0.72 $\pm$ 0.14 \\
Ridge Logistic Regression & 0.65 $\pm$ 0.12 & 0.73 $\pm$ 0.14 & 0.75 $\pm$ 0.11 & 0.73 $\pm$ 0.10 \\
AutoML & 0.69 $\pm$ 0.00 & 0.78 $\pm$ 0.00 & 0.70 $\pm$ 0.00 & 0.74 $\pm$ 0.00 \\
CatBoost & 0.66 $\pm$ 0.10 & 0.78 $\pm$ 0.10 & 0.73 $\pm$ 0.11 & 0.75 $\pm$ 0.09 \\
XGBoost & 0.66 $\pm$ 0.10 & 0.76 $\pm$ 0.10 & 0.74 $\pm$ 0.11 & 0.74 $\pm$ 0.09 \\
\bottomrule
\end{tabularx}
\end{table}

\begin{table}[H]
\centering
\caption{Experiment 4F. Synthetic augmentation to 1000 total rows after feature selection and volume inclusion. Values are mean $\pm$ SD across 100 random splits.}
\label{tab:S13}
\scriptsize
\begin{tabularx}{\textwidth}{Xcccc}
\toprule
\textbf{Model} & \textbf{Accuracy} & \textbf{Recall} & \textbf{Precision} & \textbf{F1-score} \\
\midrule
TabPFN & 0.70 $\pm$ 0.10 & 0.80 $\pm$ 0.06 & 0.77 $\pm$ 0.11 & 0.78 $\pm$ 0.08 \\
Adaptive Boosting & 0.68 $\pm$ 0.08 & 0.75 $\pm$ 0.12 & 0.77 $\pm$ 0.09 & 0.75 $\pm$ 0.09 \\
Logistic Regression & 0.63 $\pm$ 0.12 & 0.72 $\pm$ 0.14 & 0.72 $\pm$ 0.11 & 0.72 $\pm$ 0.12 \\
LightGBM & 0.68 $\pm$ 0.10 & 0.77 $\pm$ 0.13 & 0.75 $\pm$ 0.11 & 0.76 $\pm$ 0.11 \\
K-nearest neighbors & 0.60 $\pm$ 0.09 & 0.75 $\pm$ 0.15 & 0.69 $\pm$ 0.06 & 0.71 $\pm$ 0.09 \\
SVM & 0.65 $\pm$ 0.12 & 0.77 $\pm$ 0.28 & 0.65 $\pm$ 0.23 & 0.70 $\pm$ 0.25 \\
Random Forest Classifier & 0.67 $\pm$ 0.09 & 0.79 $\pm$ 0.09 & 0.73 $\pm$ 0.10 & 0.76 $\pm$ 0.08 \\
Decision Tree & 0.61 $\pm$ 0.12 & 0.68 $\pm$ 0.12 & 0.73 $\pm$ 0.13 & 0.70 $\pm$ 0.11 \\
Ridge Logistic Regression & 0.64 $\pm$ 0.12 & 0.73 $\pm$ 0.14 & 0.74 $\pm$ 0.10 & 0.73 $\pm$ 0.10 \\
AutoML & 0.44 $\pm$ 0.00 & 0.56 $\pm$ 0.00 & 0.50 $\pm$ 0.00 & 0.53 $\pm$ 0.00 \\
CatBoost & 0.67 $\pm$ 0.11 & 0.78 $\pm$ 0.10 & 0.74 $\pm$ 0.12 & 0.76 $\pm$ 0.10 \\
XGBoost & 0.67 $\pm$ 0.10 & 0.76 $\pm$ 0.10 & 0.76 $\pm$ 0.12 & 0.75 $\pm$ 0.09 \\
\bottomrule
\end{tabularx}
\end{table}

\subsection*{TabDDPM hyperparameters}
The following tables provide the hyperparameters used for the CatBoost evaluator and for the diffusion model itself.

\begin{table}[H]
\centering
\caption{TabDDPM CatBoost evaluator hyperparameters.}
\label{tab:S14_catboost_hparams}
\small
\begin{tabularx}{\textwidth}{l l X}
\toprule
\textbf{Hyperparameter} & \textbf{Value} & \textbf{Notes} \\
\midrule
\texttt{iterations} & 2000 & Max boosting rounds \\
\texttt{learning\_rate} & 0.08313 & Step size per tree \\
\texttt{depth} & 4 & Tree depth \\
\texttt{l2\_leaf\_reg} & 9.4522 & $L_2$ regularization \\
\texttt{bagging\_temperature} & 0.52185 & Bayesian sub-sampling strength \\
\texttt{leaf\_estimation\_iterations} & 5 & Newton steps per tree \\
\texttt{early\_stopping\_rounds} & 50 & Patience on validation loss \\
\texttt{od\_pval} & 0.001 & Over-fitting detector $p$-value \\

\bottomrule
\end{tabularx}
\end{table}

\begin{table}[H]
\centering
\caption{TabDDPM diffusion model and training hyperparameters.}
\label{tab:S15_ddpm_hparams}
\small
\begin{tabularx}{\textwidth}{l l X}
\toprule
\textbf{Hyperparameter} & \textbf{Value} & \textbf{Notes} \\
\midrule
\texttt{model\_type} & mlp & Backbone network \\
\texttt{d\_in (\# features)} & 10 & Input dimension (= radiomics subset) \\
\texttt{d\_layers} & 512$\times$5 $\rightarrow$ 256 & Hidden-layer widths \\
\texttt{dropout} & 0.0 & Hidden-layer dropout \\
\texttt{is\_y\_cond, num\_classes} & true, 2 & Conditional generation \\
\texttt{num\_timesteps} & 100 & Diffusion steps \\
\texttt{scheduler} & cosine & Noise schedule \\
\texttt{gaussian\_loss\_type} & mse & Denoising objective \\
\texttt{steps} & 20,000 & Optimizer update steps \\
\texttt{lr} & $2.288\times10^{-4}$ & Adam learning rate \\
\texttt{weight\_decay} & 0.0 & $L_2$ weight decay \\
\texttt{batch\_size (train / sample)} & 256 / 2000 & Minibatch sizes \\

\bottomrule
\end{tabularx}
\end{table}

\section*{TRIPOD+AI reporting map}
TRIPOD+AI is an updated reporting guideline for clinical prediction model studies that use regression or machine-learning methods \cite{tripodai}. The table below maps the present manuscript to the 27-item TRIPOD+AI checklist. It is intended as a transparent reporting aid rather than a claim that every item could be fully addressed from the available retrospective data. Items not performed, unavailable, or not applicable are stated explicitly.

{\scriptsize
\setlength{\tabcolsep}{3pt}
\renewcommand{\arraystretch}{1.18}
\begin{longtable}{>{\raggedright\arraybackslash}p{0.07\textwidth}>{\raggedright\arraybackslash}p{0.27\textwidth}>{\raggedright\arraybackslash}p{0.27\textwidth}>{\raggedright\arraybackslash}p{0.30\textwidth}}
\caption{TRIPOD+AI reporting map for the present study. The checklist item wording is summarized from TRIPOD+AI. Locations refer to the accompanying main manuscript and this Supplementary Information file.}\label{tab:S16_tripodai}\\
\toprule
\textbf{Item} & \textbf{Reporting element} & \textbf{Where reported} & \textbf{Status and comments} \\
\midrule
\endfirsthead
\caption[]{TRIPOD+AI reporting map for the present study (continued).}\\
\toprule
\textbf{Item} & \textbf{Reporting element} & \textbf{Where reported} & \textbf{Status and comments} \\
\midrule
\endhead
\midrule
\multicolumn{4}{r}{\textit{Continued on next page}}\\
\endfoot
\bottomrule
\endlastfoot
1 & Title identifies a prediction model study, target population, and predicted outcome. & Main title page. & Reported. The title identifies T2-weighted MRI radiomics, H3K27M screening, and pediatric diffuse midline glioma. \\
2 & Abstract follows TRIPOD+AI abstract principles. & Main Abstract. & Reported in unstructured Scientific Reports format. The abstract summarizes context, cohort, predictors, model components, evaluation strategy, performance, and intended screening role. \\
3a & Healthcare context and rationale, including diagnostic/prognostic context and existing models. & Main Introduction. & Reported. The manuscript explains biopsy constraints, referral imaging heterogeneity, prior radiomics/deep learning work, and the need for a T2-only screening signal. \\
3b & Target population, intended purpose, care pathway, and intended users. & Main Introduction and Discussion. & Reported. The intended population is pediatric DMG patients with available T2-weighted MRI; intended use is clinician-led triage for molecular confirmation or trial screening, not autonomous diagnosis. \\
3c & Known health inequalities between sociodemographic groups. & Main Discussion limitations; this table. & Not analyzed. Sociodemographic information sufficient for health-inequality analysis was not available in the retrospective dataset. \\
4 & Study objectives and whether development, evaluation, or both were performed. & Main Introduction and Methods. & Reported. The study develops and internally evaluates an ablation-driven T2-weighted radiomics screening framework. \\
5a & Data sources for development/evaluation and rationale/representativeness. & Main Methods: Study design and patients; MRI selection and acquisition heterogeneity. & Reported. Data came from a retrospective institutional repository with scans acquired across multiple external sources; the rationale was to reflect referral-style clinical imaging. \\
5b & Dates of collected participant data and follow-up, if applicable. & Main Methods: MRI selection and acquisition heterogeneity. & Partially reported. Imaging was collected over more than 10 years; exact accrual start and end dates should be inserted if available. Follow-up time was not part of the prediction target. \\
6a & Study setting, number and location of centers. & Main Methods and Discussion. & Reported with clarification. Scans originated from multiple institutions/scanners/protocols, but model development and evaluation were performed within one institutional repository. \\
6b & Eligibility criteria. & Main Methods: Study design and patients. & Reported. Inclusion/exclusion criteria are listed. \\
6c & Treatments received and how handled, if relevant. & Main Methods: Study design and patients. & Partially reported. Only pre-treatment MRI was included; treatment variables were not predictors and were not modeled. \\
7 & Data preprocessing and quality checking, including across sociodemographic groups if applicable. & Main Methods: MRI selection; preprocessing; radiomic feature extraction; Supplementary Methods. & Reported for imaging preprocessing. Cross-sociodemographic quality assessment was not possible because sociodemographic metadata were incomplete. \\
8a & Outcome definition, time horizon, assessment method, and consistency across groups. & Main Methods: Study design and patients. & Reported. The outcome was tissue-based H3K27M mutation status. No time horizon applies because this is a diagnostic/molecular classification task. \\
8b & Outcome assessor qualifications if subjective interpretation is required. & Main Methods; this table. & Partially reported. H3K27M status was obtained from tissue-based clinical records. Assay/pathologist-level details were not consistently available. \\
8c & Blinding of outcome assessment. & This table. & Not reported/unknown. No specific blinding procedure for molecular outcome assessment was recorded in the retrospective data. \\
9a & Choice of initial predictors and any pre-selection before model building. & Main Methods: radiomic feature extraction; feature selection; tumor volume. & Reported. Initial predictors were PyRadiomics features from original and filtered T2-weighted images, plus optional tumor volume. \\
9b & Definition and measurement of all predictors. & Main Methods; Main Table 3; Supplementary Table S5. & Reported. Radiomic feature classes, extraction settings, selected features, and volume computation are described. \\
9c & Predictor assessor qualifications if subjective interpretation is required. & Main Methods: Tumor segmentation and ROI definition. & Reported for segmentation. Tumor masks were manually delineated by a neuroradiologist with 7 years of experience; radiomic measurements were computationally extracted. \\
10 & Study size rationale and sample size calculation. & Main Methods and Discussion limitations. & Partially reported. The cohort size was determined by availability of mutation-confirmed pediatric DMG cases. No formal sample size calculation was performed; rarity and limited sample size are stated as limitations. \\
11 & Missing data handling and reasons for omitting data. & Main Methods: Study design and patients; MRI selection; Discussion limitations. & Reported at study level. Cases without molecular ground truth, diagnostic-quality pre-treatment T2-weighted MRI, or segmentation were excluded; 125 unlabeled cases were excluded from primary analysis. Incomplete scanner/protocol metadata limited subgroup analyses. \\
12a & How data were used for model development and evaluation, including partitioning. & Main Methods: Classifiers and evaluation. & Reported. Evaluation used 100 stratified 70/30 train/test splits. \\
12b & Predictor handling in analyses, including transformations or standardization. & Main Methods: Preprocessing; radiomic extraction; feature selection; synthetic augmentation. & Reported. Images were resampled, skull-stripped, filtered, converted into radiomic features, pruned by correlation, reduced by repeated RFE, and augmented only within training folds when applicable. \\
12c & Model type, rationale, model-building steps, hyperparameter tuning, and internal validation. & Main Methods; Supplementary Tables S14-S15; code repository. & Mostly reported. Classifier families, train/test splitting, leakage control, and TabDDPM hyperparameters are provided. Some classifier-specific hyperparameters should be confirmed from the repository if reviewers request full settings. \\
12d & Handling and quantification of heterogeneity across clusters. & Main Methods and Discussion limitations. & Partially addressed. Acquisition heterogeneity was intentionally retained; cluster-specific modeling or performance estimation by scanner/site was not performed because metadata were incomplete. \\
12e & Performance measures and plots, including discrimination, calibration, and clinical utility where relevant. & Main Methods, Results, figures, and tables; Supplementary Results. & Reported for accuracy, recall, precision, and F1-score. Calibration, AUC, and decision-curve/net-benefit analyses were not performed. \\
12f & Model updating after evaluation. & This table. & Not applicable. No external evaluation or model updating/recalibration was performed. \\
12g & How predictions were calculated during model evaluation. & Main Methods; code repository. & Reported in workflow and code availability. Predictions were generated by the trained classifier in each split; the repository contains the analytical code. \\
13 & Class imbalance methods and any recalibration after imbalance handling. & Main Methods: Synthetic augmentation; Main Results; Supplementary Tables S8-S15. & Reported. TabDDPM generated minority-class synthetic samples within training folds only. Recalibration after augmentation was not performed. \\
14 & Approaches used to address model fairness. & Main Discussion limitations; this table. & Not performed. Fairness analysis was not feasible because sociodemographic subgroup data were incomplete and the cohort was small. \\
15 & Model output and classification thresholds. & Main Methods: Classifiers and evaluation. & Partially reported. Models produced binary H3K27M-positive versus wild-type classifications using default classifier decision rules; no deployment threshold was optimized or proposed. \\
16 & Differences between development and evaluation data. & Main Methods: Classifiers and evaluation; Discussion limitations. & Reported. Training and test folds were drawn from the same cohort; test folds contained real patients only, whereas synthetic samples were added only to training folds. No external dataset was used. \\
17 & Ethics approval and consent/waiver. & Main Methods: Ethics and consent. & Reported with placeholder. Authors should insert the exact ethics committee name and approval number before submission. \\
18a & Funding source and funder role. & Main Acknowledgements. & Placeholder present. Funding and funder role should be completed before submission. \\
18b & Conflicts of interest. & Main Additional information. & Reported as no competing interests, pending author confirmation before submission. \\
18c & Study protocol availability. & This table. & Not prepared/available unless the authors have a protocol document. If none exists, state that no prospective protocol was prepared. \\
18d & Study registration. & This table. & Not registered unless authors have a registration record. If none exists, state that the retrospective model-development study was not registered. \\
18e & Data sharing. & Main Data Availability. & Reported. Raw imaging cannot be publicly released due to privacy and institutional restrictions; controlled access to derived de-identified features may be considered subject to approval. \\
18f & Code sharing. & Main Data Availability. & Reported. The code repository URL is provided. \\
19 & Patient and public involvement. & This table; optional statement can be added to main Additional information. & Not involved. No patient or public involvement occurred in study design, conduct, reporting, or dissemination planning. \\
20a & Participant flow and number with/without outcome. & Main Results: Cohort and imaging setting; Fig. 1. & Reported. The analyzed cohort included 98 patients, 73 H3K27M-positive and 25 wild-type. \\
20b & Participant characteristics overall and by data source/setting, including missing data. & Main Results; Methods; Discussion limitations. & Partially reported. Class distribution and imaging heterogeneity are reported. Detailed demographic, treatment, and scanner/protocol summaries were not available for reliable subgroup reporting. \\
20c & For model evaluation, comparison with development data. & Main Methods; this table. & Not applicable as an external validation item. Repeated internal splits used the same source cohort; train/test folds were stratified by mutation status. \\
21 & Number of participants and outcome events in each analysis. & Main Methods and Results. & Reported at cohort level. Repeated stratified 70/30 splits imply split-specific train/test counts; exact counts for each seed are not tabulated. \\
22 & Full model specification sufficient for prediction in new individuals and third-party evaluation. & Main Methods; Supplementary Methods; code repository. & Partially reported. Methods and code are provided for reproduction; a locked deployment model object is not provided, consistent with the exploratory screening-stage aim. \\
23a & Model performance estimates with uncertainty and subgroup results where applicable. & Main Results; Main Table 1; Supplementary Tables S1-S13. & Reported as mean $\pm$ standard deviation across 100 splits. Confidence intervals and subgroup performance were not computed. \\
23b & Heterogeneity of model performance across clusters. & Discussion limitations; this table. & Not examined because scanner/site metadata were incomplete. \\
24 & Model updating results. & This table. & Not applicable. No model updating was performed. \\
25 & Interpretation of main results, including fairness and comparison with previous studies. & Main Discussion. & Reported. The discussion interprets results as a triage/screening signal and contextualizes performance relative to prior multiparametric and external-test studies. Fairness could not be assessed. \\
26 & Limitations and effects on bias, uncertainty, and generalizability. & Main Discussion. & Reported. Limitations include retrospective single-repository development/evaluation, no external validation, moderate performance, T2-only input, small rare-disease cohort, incomplete metadata, and feature-space augmentation. \\
27a & Handling poor-quality or unavailable input data in implementation. & Main Methods and Discussion. & Partially reported. Non-diagnostic images were excluded. Future implementation would require diagnostic-quality T2-weighted imaging and segmentation quality control. \\
27b & User interaction and expertise required. & Main Methods and Discussion. & Reported. Current workflow requires expert tumor segmentation and clinician interpretation; the tool is positioned as human-in-the-loop decision support. \\
27c & Next steps for applicability and generalizability. & Main Discussion. & Reported. Future work should include external validation, calibration, decision-analytic evaluation, and broader subgroup assessment when metadata are available. \\
\end{longtable}
}